\documentstyle[12pt,aaspp]{article}

  at 12.0 true pt 

\def\etal{{et al}}

\def\phx{{\tt PHOENIX}}

\def\water{{H$_2$O}}

\def\b{\beta}

\def\div#1#2{{#1\over #2}}
\def\rout{\ifmmode{r_{\rm out}}\else\hbox{$r_{\rm out}$}\fi}
\def\tmax{\ifmmode{\tau_{\rm max}}\else\hbox{$\tau_{\rm max}$}\fi}
\def\tstd{\ifmmode{\tau_{\rm std}}\else\hbox{$\tau_{\rm std}$}\fi}
\def\vmax{\ifmmode{v_{\rm max}}\else\hbox{$v_{\rm max}$}\fi}
\def\muE{\ifmmode{\mu_{\rm E}}\else\hbox{$\mu_{\rm E}$}\fi} 
\def\pE{\ifmmode{p_{\rm E}}\else\hbox{$p_{\rm E}$}\fi} 
\def\bmax{\ifmmode{\b_{\rm max}}\else\hbox{$\b_{\rm max}$}\fi}
\def\kms{\hbox{$\,$km$\,$s$^{-1}$}}

\def\ang{\hbox{\AA}}

\def\Teff{\hbox{$\,T_{\rm eff}$} }
\def\alog#1{\times 10^{#1}}

\def\rout{\hbox{$r_{\rm out}$} }
\def\chistd{\ifmmode{\chi_{\rm std}}\else\hbox{$\chi_{\rm std}$}\fi}

\def\lstar{\ifmmode{\Lambda^*}\else\hbox{$\Lambda^*$}\fi} 
\def\Rop{\ifmmode{[R_{ij}]}\else\hbox{$[R_{ij}]$}\fi}

\def\Rji{\ifmmode{[R_{ji}]}\else\hbox{$[R_{ji}]$}\fi}
\def\Rstar{\ifmmode{[R_{ij}^*]}\else\hbox{$[R_{ij}^*]$}\fi}

\def\Rjistar{\ifmmode{[R_{ji}^*]}\else\hbox{$[R_{ji}^*]$}\fi}
\def\DRji{\ifmmode{[\Delta R_{ji}]}\else\hbox{$[\Delta R_{ji}]$}\fi}
\def\DRij{\ifmmode{[\Delta R_{ij}]}\else\hbox{$[\Delta R_{ij}]$}\fi}

\def\ns{\ifmmode{N_{\rm s}}          
        \else\hbox{$N_{\rm s}$}\fi}

\def\mat#1{{\bf #1}}     
\def\vek#1{{#1}}         

\newcount\eqcount
\def
  \nummer{
    \global\advance\eqcount by 1
    (\the\eqcount)
  }

\def
  \numadv{
    \global\advance\eqcount by 1
  }

\def
   \numout#1{
     (\the\eqcount #1)
  }

\def\ivek#1#2{\ifmmode{\vek{I}^{#1}_{#2}}
        \else\hbox{$\vek{I}^{#1}_{#2}$}\fi}

\def\tmat#1#2{\ifmmode{\mat{t}^{#1}_{#2}}
        \else\hbox{$\mat{t}^{#1}_{#2}$}\fi}
\def\rmat#1#2{\ifmmode{\mat{r}^{#1}_{#2}}
        \else\hbox{$\mat{r}^{#1}_{#2}$}\fi}
\def\bvek#1#2{\ifmmode{\beta^{#1}_{#2}}
        \else\hbox{$\beta^{#1}_{#2}$}\fi}

\def\lp{\ifmmode{\lambda^+_\tau}           
        \else\hbox{$\lambda^+_\tau$}\fi}
\def\lm{\ifmmode\lambda^-_\tau             
        \else\hbox{$\lambda^-_\tau$}\fi}
\def\la{\mathrel{\hbox{\rlap{\hbox{\lower4pt\hbox{$\sim$}}}\hbox{$<$}}}}
\def\ga{\mathrel{\hbox{\rlap{\hbox{\lower4pt\hbox{$\sim$}}}\hbox{$>$}}}}

\def\be{\begin{eqnarray}}
\def\ee{\end{eqnarray}}

\def\etal{et al.}

\def\k{\,{\rm K}}

\def\teff{{\rm T}_{\rm eff}}

\def\eps{\epsilon}

\begin{document}
\baselineskip=12pt
\bibliographystyle{apj}

\title{Model Atmospheres for M (Sub)Dwarf stars\\
I. The base model grid}

\author{France Allard}
\affil{Dept.\ of Geophysics and Astronomy, University of British Columbia, \\
Vancouver, B.C., V6T 1Z4 Canada\\
E-Mail: allard@astro.ubc.ca}

\and

\author{Peter H. Hauschildt}
\affil{Dept.\ of Physics and Astronomy, Arizona State University, Box 871504,\\ 
Tempe, AZ 85287-1504\\
E-Mail: yeti@sara.la.asu.edu}

\begin{abstract}

We have calculated a grid of more than 700 model atmospheres valid for 
a wide range of parameters encompassing the coolest known M~dwarfs, 
M~subdwarfs and brown dwarf candidates: $1500\le \teff \le 4000\,$K, 
$3.5\le \log(g)\le 5.5$, and $-4.0\le { [M/H]}\le +0.5$. 

Our equation of state includes 105 molecules and up to 27 
ionization stages of 39 elements.  In the calculations of the base 
grid of model atmospheres presented here, we include over 300 molecular 
bands of 4 molecules (TiO, VO, CaH, FeH) in the JOLA approximation, 
the water opacity of Ludwig (1971), collision induced opacities,
b-f and f-f atomic processes, as well as about 2 million spectral 
lines selected from a list with more than 42 million atomic and 24 
million molecular (H$_2$, CH, NH, OH, MgH, SiH, C$_2$, CN, CO, SiO) 
lines. High-resolution synthetic spectra are obtained using an opacity 
sampling method. The model atmospheres and spectra are calculated with 
the generalized stellar atmosphere code \phx, assuming LTE, plane-parallel 
geometry, energy (radiative plus convective) conservation, and hydrostatic 
equilibrium.  

The model spectra give close agreement with observations of M~dwarfs 
across a wide spectral range from the blue to the near-IR, with one 
notable exception: the fit to the water bands.  We discuss several 
practical applications of our model grid, e.g., broadband colors 
derived from the synthetic spectra.  In light of current efforts to 
identify genuine brown dwarfs, we also show how low-resolution spectra 
of cool dwarfs vary with surface gravity, and how the high-resolution 
line profile of the Li~I resonance doublet depends on the Li abundance.

\end{abstract}

\section{Introduction}

Interest in the extreme lower main sequence has blossomed in recent 
years, largely because of its implications for star formation theory 
and dark matter candidates.  Advances in spectroscopy and photometry, 
especially at infrared wavelengths, promise a wealth of new data on 
low-mass stars.  Therefore, parallel improvements in models of cool 
stellar atmospheres are very important if the full potential of these 
data is to be realized.

The lower main sequence is populated by a variety of objects --- old 
disc M~dwarfs, halo population subdwarfs, and possibly even young,
still contracting, brown dwarfs.  The proper detection and classification 
of these objects requires synthetic spectra which can quantify their basic 
properties: elemental abundances, effective temperature, surface gravity,
and (if the parallax is available) the radius, mass, and luminosity. 
At present, these physical properties are not yet particularly well 
determined.  Traditional techniques to estimate stellar effective 
temperature based on blackbody approximations and broadband photometry 
are at best dangerous (and at worst, invalid) for cool M dwarfs, whose 
true continua are masked by extensive molecular absorption.  Therefore, 
``observed'' positions of these stars in the HR Diagram are highly uncertain.
Such uncertainties can have profound implications for star formation theory 
and the search for brown dwarfs (Burrows and Liebert, 1993)\nocite{hubb94}.

The success of modeling atmospheres for low-mass stars relies heavily 
on the treatment and quality of the opacities. The first grid of model 
atmospheres of M~dwarfs to include molecular and atomic line blanketing 
as well as convection by Mould (1976)\nocite{mould76}, based on his thesis, 
covered an effective temperature range from 4250$\,$K down to 3000$\,$K, 
whereas many of the faint well-observed stars have colors implying 
temperatures as much as 1200$\,$K cooler (H. Jones, private communication).  
Although Mould's models represented an important breakthrough to the lower 
main sequence, they nevertheless failed to reproduce key observed 
characteristics of M~dwarfs even in the range of $\teff$ they covered.

Since the early computations of Mould, significant improvements to
molecular and atomic data bases have been achieved, enabling the 
construction of model atmosphere grids for M dwarfs which can be 
compared directly to the coolest observed spectra. A new generation 
of model calculations, by Allard (1990)\nocite{allard90} and Kui 
(1991)\nocite{kui91}, was able to reproduce many of the observed 
features found in the optical and near-IR.  The first set of Allard 
model atmospheres were used by Kirkpatrick \etal\ (1993)\nocite{kirk93} 
to obtain an M~dwarf temperature sequence which succeeded in bringing 
at least the hotter dM's close to the theoretical main sequence. 

However, like Mould's models before them, these models disagreed with 
observations on several points. For example, they did not reproduce 
$I-J$ colors of dM stars, they could not match the IR flux distribution 
dominated by water band absorption, and significant problems remained 
in the UV and IR for the low metallicity subdwarfs as well.

Allard (1993)\nocite{faIAU146} has addressed and corrected several of 
these discrepancies in a test model of the spectral distribution of 
the classical dwarf. In this paper, we present an improved generation of
models which incorporates the results of Allard (1993). We also include 
many more metal lines, which are critical for the energy balance and 
serve as important diagnostics of the synthetic spectra. The numbers of 
molecules in both the equation of state (EOS) and the opacity calculations 
has been significantly increased compared to any previous calculations.  
Furthermore, we have extended our opacity sampling method to include up 
to 25 million molecular lines, which both replace older Just Overlapping 
Line Approximation (JOLA) band opacities and added new opacity sources. 
This new grid of cool dwarf atmospheres significantly improves upon the 
work of Allard (1990)\nocite{allard90}.

In the next section we present the most important aspects of the model 
construction and discuss the model grid and the effects of some model 
parameters on the model structure and on the resulting synthetic spectra  
in section \ref{grid}. Finally we summarize the main properties of the 
new grid in section \ref{colors} by confronting it with observations in 
selected color-color planes. 

\section{Model construction}

Our model atmospheres are constructed under the classical assumptions 
of stellar atmosphere theory, implemented in our model atmosphere code 
\phx, version 4.9. \phx\ was first developed to treat the expanding 
ejecta of novae and supernovae; as a result, it is capable of solving 
the (special relativistic) radiative transfer equation (RTE) in the 
Lagrangian frame self-consistently with multi-level, non-LTE rate 
equations and the (special relativistic) radiative equilibrium (RE) 
equation in the Lagrangian frame. Numerical methods used in \phx\ 
include the solution of: (i) the RTE using the Accelerated Lambda 
Iteration (ALI) method described by Hauschildt (1992)\nocite{s3pap} 
and Hauschildt \etal\ (1994c)\nocite{aliperf}, and the (ii) multi-level 
non-LTE continuum and line transfer problem using the ALI algorithm 
(Hauschildt, 1993)\nocite{casspap}. More details on these aspects of 
the code can be found in Hauschildt \etal\ (1994b)\nocite{cygpap} 
and Baron, Hauschildt \& Branch (1994)\nocite{sn93jpap}.

Important additions to \phx\ were necessary to handle the problems 
specific to compact cool dwarf atmospheres. The most important 
modifications, many of which were adapted from Allard (1990, 1993), 
are: (i) solution of the equation of state (EOS) simultaneously for 
more than 334 species including diatomic and polyatomic molecules, (ii) 
inclusion of Van der Waals atomic line broadening, (iii) inclusion 
of molecular opacities in the Just Overlapping Line Approximation 
(JOLA) as well as several pre-tabulated Straight Mean (SM) molecular 
opacities, (iv) inclusion of molecular lines for a number of molecules, 
(v) treatment of convection using the mixing length formalism, and 
(vi) solution of the RE equation by a modified Uns\"old-Lucy method. 

The model atmosphere problem (i.e., the self-consistent simultaneous 
solution of the EOS, the hydrostatic equation, the RTE at all 
wavelengths, the radiative-plus-convective energy equation, and 
the NLTE multi-level transfer and rate equations) is treated 
using a nested iteration scheme like that described by Hauschildt 
(1991)\nocite{yetiphd}.  

Not all aspects of cool dwarf stars lead to additional complications 
in the model calculations; some actually allow us simplifications that 
substantially reduce the CPU time required per model calculation.  
First, since their surface gravities are fairly high ($\log(g) > 3$) 
such that the atmospheres have a relative extension of less than 1\%, 
we can safely treat those atmospheres using a plane-parallel and static 
approximation.  We have performed test calculations with the spherically 
symmetric mode of \phx\ for some models with $\log(g)=3.0$ and found (as
expected) no significant differences in the models and synthetic spectra 
compared to the plane-parallel case. Second, the velocities of the 
convective cells in these stars are too small to be detectable in the 
low-resolution spectra we wish to reproduce, and furthermore, have a 
negligible effect on the transfer of line radiation.  Therefore, we 
follow Allard's (1990) approach and neglect the effects of convective 
motions on line formation. Finally, we neglect for simplicity possible 
departures from LTE and external radiation fields.

The model atmospheres are characterized by the following parameters:
(i) the surface gravity, $\log(g)$, (ii) the effective temperature, 
$\teff\unskip$, (iii) the mixing length to scale height ratio, $l$, 
here taken to be unity, (iv) the isotropic micro-turbulent velocity , 
$\xi$, and (v) the element abundances, here parameterized by $[M/H]$, 
the logarithm of the metal abundance (by number) relative to the solar 
abundances (\cite{solab89}).

\subsection{Hydrostatic equation and optical depth grid}

\phx\ uses a prescribed optical depth grid to perform the model 
calculations. This grid is defined at one chosen wavelength.
For this paper, we take the ``standard optical depth'' $\tstd$
as the continuum optical depth in absorption at $1.2\,\mu$m. We 
use a logarithmic $\tstd$ grid.  A representative grid typically 
contains 50 points between $\tstd=10^{-10}$ and $10^2$, which 
ensures good $\tau$ resolution for all wavelength points in the 
model. This is important since opacities in M dwarfs are extremely 
non-grey due to the simultaneous presence of lines, bands and 
continuum. This particular $\tstd$ grid was also chosen such as
to assure both high accuracy and high 
performance of the ALI radiative transfer 
method used in \phx.

To solve the hydrostatic equation,
\[
      \div{dP}{dr} = -g\rho
\]
where $P=P_g+P_r$ is the total (gas pressure $P_g$ and radiative 
pressure $P_r$) pressure, r the radius, $g$ the gravity, and $\rho$ 
the density, we first introduce the standard optical depth $\tstd$ 
as a new independent variable via $d\tstd = -\chistd dr$, with the 
initial condition $\tstd=0$ for $r=\rout$.  Hence, we obtain
\[
     \div{dP}{d\tstd} = \div{g\rho}{\chistd}.
\]
At the beginning of each iteration, we integrate this equation 
numerically for the current run of the electron temperature 
$T_e(\tstd)$ in the radiative zones and simultaneously with 
the adiabatic temperature gradient equation for the convective 
zones using a Bulirsch-Stoer numerical quadrature method with 
adaptive step size; see Allard (1990)\nocite{allard90} for 
details of this procedure.  For the conditions present in M dwarfs, 
the radiation pressure gradient as well as the gradient of the 
turbulent pressure are much smaller than the gas pressure in all 
models considered here, so that we can set $P=P_g$ without loss 
of accuracy.

\subsection{EOS and adiabatic gradient}

The equation of state (EOS) used in this work includes up to 26
ionization stages of 39 elements (H, {He}, Li, Be, B, {C}, {N},
{O}, F, Ne, Na, Mg, Al, Si, P, S, Cl, Ar, K, Ca, Sc, Ti, V, Cr, 
Mn, Fe, Co, Ni, Cu, Zn, Ga, Kr, Rb, Sr, Y, Zr, Nb, Ba, and La) 
as well as 105 molecules (\cite{allard90,faIAU146}, see Table~1 
for a list of the most important molecules). Although grains and 
dust may be of importance in the atmospheres of the coolest M~dwarfs, 
they have not been included at this level of our calculations but 
are planned for future studies. 

We assume particle and charge conservation in each layer of an ideal 
gas. Although \phx\ generally accounts for non-LTE effects of important 
species (H, He, Na~I, Mg~II,  Ca~II, Ne~I, and Fe~II), global LTE conditions 
are assumed here for simplicity.  We solve the EOS using a complete 
linearization scheme treating {\em all} considered species.  Our method 
proved highly stable (even under extreme conditions of low temperature 
and high pressure), while being reliable and yet flexible to changes in 
elemental abundances and chemical composition.  However, it is relatively 
slow on non-vectorizing computers. Therefore, we have compiled partial 
pressure tables which the model calculations may use. These tables extend 
over typical parameter ranges of:
$$ 0.3 \le \theta = 5040/T \le 5.25 \qquad ; \qquad \Delta{\theta}=0.05$$
$$ -4.00 \le \log P_g \le 12.00   \qquad ; \qquad \Delta{\log P_g}=0.20$$
for each mixture covered by our grid of model atmospheres.

The full composition of 334 species (listed in Table~1 for its most  
relevant section) was selected for the table construction so as to 
obtain accurate partial pressures for the species involved in opacity 
calculations. These include ${\rm H}_2$, ${\rm H}_2$O, CN, TiO, OH, CH, 
CO, VO, MgH, SiH, CaH, and FeH.  The EOS tables are used {\em only} 
during the solution of the hydrostatic equation. Once this part of a 
model iteration is completed, the full EOS is solved with all species 
at all depth points in order to (a) check the accuracy of the solution 
and (b) ensure that the partial pressures of {\em all} species are known 
accurately at all depth points.  The EOS tables are used, therefore, 
only to speed up the numerical calculations. We have calculated some 
of the models without using the EOS tables and found no significant 
change. 

Our partition functions of atoms and ions are calculated internally 
by summing over the bound states taken from Mihalas (1993, private 
communication).  The partition functions of most diatomic molecules 
are calculated according to the Tatum (1966)\nocite{tatum66} approximation.  
In the case of H$_2$ and CO, however, we opted for the more accurate 
polynomial expression by Irwin (1987)\nocite{irwin87}. The partition 
functions of polyatomic molecules are also taken from Irwin 
(1988)\nocite{irwin88}.

The dissociation constants of diatomic and polyatomic molecules are 
calculated internally by means of a Saha-like derivation (Tatum, 
1966)\nocite{tatum66}. Atomization energies (energy relative to the
neutral monoatomic reference species) are taken from the literature 
(e.g. CN with 7.77$\,$eV from Costes, Naulin \& Dorthe, 1990\nocite{costes90};
FeH with 1.63$\,$eV from Schultz \& Armentrout, 1991)\nocite{schultz91}, 
and from the compilations by Irwin (1988), Huber \& Herzberg 
(1979)\nocite{hh79} and Rosen (1970)\nocite{rosen70}.

The adiabatic gradient is calculated {\em analytically} including 
the most important species; H~I--II, He~I--II, H$_2$, CO, and 
electrons (as well as a number of species of secondary importance, 
for a total of 12 species); based on an algorithm developed by Wehrse 
(1977)\nocite{wehrse77}. Details of the numerical application of this 
algorithm may also be found in Allard (1990)\nocite{allard90}.  This 
method has the advantage of high speed compared to numerical differentiation 
methods while giving very accurate results. Additional species can be 
easily included if required by, for instance, a different mixture. The 
results of the adiabatic gradient calculations are in very good agreement 
with the results obtained through numerical differentiation of the entropy 
by Uns\"old (1968)\nocite{unsoeld55} and by Wehrse (1974)\nocite{wehrse74}.

\subsection{Continuous opacities}

The conventional continuous opacity sources included in our model 
calculations are ${\rm H}^-$ (John 1988)\nocite{john88}, ${\rm He}^-$ 
and ${\rm H}_2^-$ (Vardya 1966)\nocite{vardya66}, the ${\rm H}_2^+$ 
and quasi-molecular hydrogen opacities (Carbon \& Gingerich 1969; 
Gingerich 1971)\nocite{carbging69,ging71}, ${\rm C}^-$ opacities 
(McDowell \& Myerscough 1966)\nocite{mcdowell66}. The most important 
b-f and f-f atomic processes (\cite{cygpap,faIAU146}) are also included, 
based on cross-sections compiled by Mathisen (1984)\nocite{mathisen84}.

However, under the atmospheric conditions involved in this study,
molecular opacities dominate by several orders of magnitudes the
conventional continuous opacities all across the entire spectrum.
In the metal-rich regime, water vapor dominates the IR opacity.
The H$_2$O absorption coefficients are therefore considered as a
continuous opacity source in our models. We used for this purpose
the Ludwig's (1971)\nocite{ludwig} tables of straight means opacities. 
This choice is motivated by the fact that the only available list 
of \water\ transitions, although describing the infrared bands well, 
fails to reproduce the continuum around $1.2\,\mu$m due to its 
increasing incompleteness toward J-numbers 
greater than 30 (see also \cite{h2o94}). The Ludwig data seems on 
the other hand to provide a satisfactorily description of the tip 
of the energy distribution (Allard 1993).  The Ludwig data are valid 
for $300\k \le {\rm T}_e \le 3000\k$. Since electronic temperatures 
within the model photospheres often exceed 3000\k, these SM opacity 
profile have been conservatively extrapolated up to 4000~K 
(\cite{faIAU146}). In spectral ranges not covered by the Ludwig's 
tables ($\lambda < 1\mu$m), we have adopted JOLA \water\ opacities 
(Tsuji 1966)\nocite{tsuji66}.

Similarly, in metal-poor regimes where pressure-induced H$_2$ 
opacities become the leading IR opacity source, we have included 
the Collision-Induced Absorptions (CIA, Borysow 1993 and references 
therein) of ${\rm H}_2$-${\rm H}_2$, ${\rm H}_2$-H and ${\rm H}_2$-He 
based on Lenzuni \etal\ (1991) as continuous opacity sources in our 
models.

Fig.~\ref{opspc} (a) and (b) the resulting continuous opacity 
profiles for these two opacity limits, i.e. a solar and an extreme metal-poor 
([M/H]=-2.5) mixture, where the gas conditions are chosen to represent
typical model conditions.

\subsection{Molecular Band Opacities}

As pointed out in the previous section, molecular band absorption 
is the primary source of blanketing in most spectral regions of 
M~dwarf atmospheres.  The millions of molecular transitions, however,  
often lack even approximate oscillator strengths, while the broadening 
mechanisms in the high gas pressure conditions encountered in M~dwarf 
atmospheres are also rather poorly understood. This situation renders 
a line-by-line approach to the molecular opacities of M~dwarfs far 
more complicated than in the case of extended M giant atmospheres.  
But precisely because of the higher pressure conditions, M~dwarfs 
are presumed better suited to a statistical treatment of their 
molecular opacities.  The band model techniques, for example, offer 
an interesting alternative for the dense atmospheres of cool dwarfs.  

For this work we use Golden's (1967)\nocite{golden67} formalism of 
the JOLA technique. For reversed bands we used also the more general 
JOLA formulation by Zeidler \& Koester (1982)\nocite{zk82}. 
This technique allows to approximate the absorption within a band 
with help of only a limited number of molecular constants: the 
rotational line structure in the bands is then reproduced by a 
continuum distribution.  These molecular constants are generally 
taken from Huber \& Herzberg (1979)\nocite{hh79} or alternatively 
from Rosen (1970)\nocite{rosen70}.  

The applicability of this technique depends on the typical line 
spacing in each given molecular band; i.e., it assumes that the 
molecular lines overlap without being saturated (see also Tsuji 
1994\nocite{tsuji94} for a review of the subject).  Although the lines of some 
important molecules such as \water\ are expected to overlap, this 
is certainly not the case for several molecules of importance in 
cool dwarfs spectra (see \cite{kui91}). Therefore, we have used 
the JOLA approach only for molecular bands for which we lack all 
necessary line data.

For instance, the rotational analysis by Phillips \etal\ (1987)
\nocite{phillips87} has allowed a major improvement to the model 
spectra of Allard (1993): the inclusion of FeH opacities. The most 
striking FeH feature is the Wing-Ford band, associated to the 
$\Delta{\nu}=0$ transitions of the $F\Delta^4-X\Delta^4$ system 
of the molecule. Those FeH features are seen in all M~dwarfs of 
spectral type later than M6 and their strengths are observed to 
increase significantly toward lower effective temperatures. This 
makes of FeH an attractive parameter indicator for cool dwarfs and 
brown dwarfs candidates close to the stellar mass limit. 

For the opacities of the TiO molecules, which define the visual 
and near-IR continuum in M~dwarfs, we prefer the tabulated 
Straight Mean (SM) coefficients by Collins (1975)\nocite{coll75}.  However, 
we had to resort to the JOLA approach for two of the TiO systems 
which where not included in Collins' original analysis: $\epsilon$
and $\phi$.

Also included in the JOLA approach are the near-IR bands of CaH 
and VO.  All other molecular absorbers are included in a line-by-line 
treatment as described in section \ref{mollines} below. However, several 
bands of interest could not be included by lack of even the most 
elementary spectroscopic data needed to apply the JOLA (e.g CaOH). 
\nocite{davis86} 

\subsection{Oscillator strengths for JOLA bands}

For explorers in the realm of molecular astrophysics, the most 
valuable -- and rarest -- treasures are accurate oscillator strengths 
for molecular bands.  Under ideal circumstances, we prefer laboratory
values when they exist, but in several cases we must resort to deriving 
``astrophysical'' or empirical values. This is done iteratively by 
fitting the observed strengths of corresponding features observed 
in the spectra of an actual star.

Once the electronic band oscillator strength is available, we 
deduce the vibrational oscillator strengths for each subbands 
and branch using the relation by Golden (1967)\nocite{golden67}: 
$$ f_{{\nu}'{\nu}''} = f_{el} \, q_{{\nu}'{\nu}''} \, 
{\omega_{{\nu}'{\nu}''} \over \omega_0}  $$
The Franck-Condon factors $q_{{\nu}'{\nu}''}$, when unavailable, 
are calculated according to Nicholls 
(1981, 1982)\nocite{nicholls81,nicholls82}. 

Among the bands considered in the JOLA approach, only TiO gains 
of laboratory band oscillator strengths. For these and all the 
bands included in the SM of collins we use the values by Davis 
\etal (1986).  

Astrophysical values are used for the oscillator strengths of the
CaH, FeH, VO, and for the TiO $\epsilon$ systems which were not 
included in Collins' (1975) work.  Allard (1993) already derived 
astrophysical oscillator strengths for many of the JOLA bands of 
the present work in a test model calculation based on the spectrum 
of the classical metal-rich M~dwarf VB~10. These values were updated
for the new conditions introduced in the present models by the 
inclusion of new line opacities.  Also, although the most important 
bands of the FeH $F\Delta^4-X\Delta^4$ system were already included
in the 1993 models, the low resolution of the observed spectrum of 
VB~10 prevented at that time the inclusion of the complex subband 
structure.  The spectra did not resolve the subband structure, 
making it impossible to determine the repartitions of oscillator 
strengths at the subband structure level. In this work we include 
the subband structure, based on higher resolution spectra of VB~10 
by Jones (private communication).  The SM and JOLA opacity data and 
the adopted values of the electronic band oscillator strengths 
are summarized in Table~2.

Fig.~\ref{VB10} presents the overall fit to the energy distribution 
of VB~10 obtained with this approach. Beside the apparent discrepancies
in the IR (discussed below), the quality of this fit is a significant
improvement which eliminates most of the problems reported by 
\cite{kirk93}.  We also verified that the strengths of the JOLA 
bands provide equally good fits to corresponding features observed 
in M~dwarfs all along the Kirkpatrick \etal\ (1993) temperature 
sequence. This bolsters our confidence in using these values 
throughout the present grid.   

\subsection{Line lists and line treatment\label{mollines}}

The \phx\ line database includes at present close to 70 million 
transitions: $\sim$~42 million atomic and ionic lines, $\sim$22 
million diatomic molecular lines (Kurucz 1993a,b)\nocite{cdrom1,cdrom15}, 
and $\sim$ 6.2 million \water\ (\cite{h2olist}). Among the molecular
species represented in \phx\'s database are: H$_2$, CH, NH, OH, MgH, 
SiH, C$_2$, CN, CO, SiO and their isotopic equivalents. 

For this work, as mentioned earlier, we adopt the Ludwig straight 
mean opacities for water. Therefore, the \water\ line list is not 
included in the model grid calculations, but is used specifically 
for high-resolution IR spectroscopic studies (e.g. Jones \etal, 
in preparation). 

Not all of the line transitions need to be considered for every 
T,$\rho$ point of an atmosphere.  Therefore, before the calculation 
of the radiation field, a smaller subset is formed from the original 
lists. First, three representative points are chosen in the atmosphere. 
Then using the densities and temperatures for these points, the 
absorption coefficient in the line center, $\kappa_l$, is calculated 
for {\em every} line and compared to the corresponding continuum 
absorption coefficient, $\kappa_c$.  A line is transferred to the 
``short list'' if the ratio $\Gamma\equiv \kappa_l / \kappa_c $ is 
larger than a pre-specified value of $ \Gamma $ (usually $10^{-4}$) 
in at least one of the test points.  In the subsequent  
calculations, all lines selected in this way are treated as individual 
lines; all others from the large line list are neglected.  Using this 
selection process, we typically find that $\sim 5\alog{4}$ (low T) to 
$2\alog{6}$ (high T) metal lines, $\sim 6\alog{6}$ molecular lines, 
and $\sim 2.1\alog{6}$ \water\ lines (when used in place of the 
Ludwig data) are stronger than about $10^{-4}$ of the local continuum 
absorption  (see \cite{h2olet}).  

We have tested this procedure by including {\em all} available molecular
lines in a few models and found no significant differences in either the 
model structure or the emergent spectrum. However, we found that larger 
values of $\Gamma$ do {\em not} include a sufficient number of molecular
lines; therefore, we use in all model calculations a value of 
$\Gamma=10^{-4}$.  A corresponding test of our line selection method for 
the atomic lines is given in Hauschildt \etal\ (1994b)\nocite{cygpap}.
We want to caution the reader that it is not clear how complete the 
line lists are for the conditions prevailing in M dwarf atmospheres.
It is possible that many transitions -- particularly in the infrared -- 
may be missing from the lists and hence from our models. 

For the calculation of the model atmospheres, we use a direct opacity 
sampling method on a fine wavelength grid with about 20,000 wavelength 
points. A typical model calculation requires about 10 model iterations 
to reach an accuracy of better than 1\%\ for both radiative equilibrium 
conditions (flux constancy and $\int \chi(S-J)\,d\lambda=0$) in the 
layers that are in radiative equilibrium and temperature changes of less 
than $1\,$K in the convective layers.  This is the case when only 
relatively poor initial estimates for the model structure are available; 
only 3 to 5 model iterations are required when we have better starting 
values. On an {\tt IBM RS/6000-580} workstation, the CPU time required 
for 10 model iterations is about $90\,$min and about $100\,$MB of main 
memory are required (mostly for the storage of the line lists). On a 
Cray C90 computer, the CPU times are typically reduced by factors of 
5--10 depending on the model to be calculated.

\phx\ includes various options for the treatment of line broadening. 
For the purpose of this model grid and low-resolution synthetic 
spectra, we use simple Gaussian profiles in order to save computer 
time.  Calculations using full pressure broadening and 
Voigt profiles are typically a factor of 3 to 5 slower. For this 
reason, we include such a treatment only in the generation of the 
final high-resolution spectra, where care is taken to reiterate 
the models in cases where the model structure proves sensitive to 
the treatment of line profiles. The Voigt profiles include Van der 
Waals interactions assuming that the collisional constants for the 
most numerous perturbers (i.e. ${\rm H}_2$ and He$\,$I in M~(sub)dwarfs 
atmospheres) do not differ significantly from their value for 
perturbations due to neutral hydrogen (Allard 1990).  We prefer
the Van der Waals damping constant calculated from the classical 
Uns\"old (1968)\nocite{unsoeld55} $C_6$ value, which we increment 
by $\Delta\log C_6=1.8$ as suggested by Wehrse \& Liebert (1980)
\nocite{wehrselieb80}, to the damping constants provided with the
Kurucz list. Potential differences between the two approaches are 
being investigated and will be reported elsewhere.

A component of thermal Doppler broadening is also included which is 
often negligible in the line wings compared to pressure broadening, 
but dominates the line cores for the microturbulent velocities 
considered in our models ($\xi=2\kms$).

\section{The base model grid\label{grid}}

Using the methods described in the previous section, we have computed a 
``base'' grid of about 700 model atmospheres. The model grid currently 
covers a wide range of parameters encompassing the coolest known M dwarfs, 
M subdwarfs and brown dwarf candidates:  $1500\le \teff \le 4000\,$K 
(with a typical stepsize of 100--$500\,$K), $3.5\le \log(g)\le 5.5$
($\Delta\log(g)=0.5$), and $-4.0\le [M/H] \le +0.5$ ($\Delta [M/H]=0.5$).
The relative 
abundances of the isotopes of each element are assumed to be solar.

 By ``base'' we mean here a grid calculated using opacity sampling for 
the atomic and molecular lines, but using Straight Mean opacities for
the dominant absorbers (TiO and \water) and the JOLA technique for 
the remaining molecular opacities.  We have several motivations and 
justifications for these assumptions.  (i) This is an necessary and 
unavoidable first step to demonstrate that the Straight Mean and JOLA 
techniques are valid under the conditions prevailing in cool star 
atmospheres. (ii) The lists of molecular transitions of several 
important absorbers including \water\ (\cite{h2o94}), if available at 
all, are still incomplete for the higher temperature conditions of 
dwarf star plasmas. (iii) The models can be used as starting point 
for a grid computed using a line-by-line treatment of the dominant 
opacities due to the TiO and \water\ lines (Allard \& Hauschildt, 
in preparation). 

We are currently extending and refining this grid as well as 
recalculating selected models with improved treatment of the molecular 
opacities. It is our intention to make the model structures plus 
the synthetic spectra available to the community. Because of the large  
number of combinations of input parameters for various synthetic spectra, 
we will not produce high-resolution synthetic spectra for large wavelength 
intervals; rather, we are prepared to calculate specific segments of 
specialized synthetic spectra upon request. 

\subsection{The structure of the model atmospheres}

In Figs.~\ref{struc1} and \ref{struc2} we display the structures of a 
small but representative subset of the model grid for a cool model 
($\Teff = 1600\,$K) and 4 hotter models ($\Teff \ge 2500\,$K).  Here 
we adopt two values of the metallicity -- $[M/H]=0$ (solar, full curves) 
and $[M/H]=-4$ (dotted curves) -- and representative values of the 
surface gravity -- $\log(g)=3.5$, 4.0, 5.0, and 5.5 for $\Teff = 
1600\,$K and $\log(g)=4.0$, 5.0, and 5.5 for $\Teff\ge 2500\,$K. 
In general, the models with the smaller $\log(g)$ also have smaller 
pressures in the inner layers of their atmospheres. The structures of 
the other models are similar and are not shown here so as not to 
clutter the figures. 

The figures show that the lower metallicity models have much higher gas 
pressures than the models with solar abundances, but that the electron 
pressures of the low $[M/H]$ models are significantly smaller. In 
addition, the lower metallicity models show a temperature inversion zone in 
the outer, optically thin ($\tstd \le 10^{-4}$) layers of the atmosphere. 
This is caused by an increase in the opacity induced by methane: Since the 
formation of methane (C${\rm H}_4$) is enhanced at the expense of CO due to 
the low-temperature and high-pressure conditions at lower metallicities,
the resulting amount of free oxygen is bound in ${\rm H}_2$O, TiO and VO
leading to a relative increase of the mean molecular opacity near
$\tstd=10^{-4}$. This effect is not only seen at very low metallicity (as
seen in Fig.~\ref{struc1}) but is even observed at solar metallicity in 
very cool models (${\rm T}_{\rm eff}\le 2000\,$K) of high gravity 
($\log g \ge 6.0$); see Allard (1990)\nocite{allard90}. 

The run of the electron pressure with temperature can be very complicated 
in the outer optically thin parts of the atmosphere due to the rapid change of
the gas pressure with a comparatively slow change in the electron temperature,
plus the presence of a temperature inversion zone. However, in the 
line-forming regions of the atmosphere, the run of $P_e$ vs. $\teff$ is 
usually very smooth. Particularly in the cooler models, we find that 
the electron pressures are $>6\,$dex smaller than the gas pressures and 
practically all of the electrons are supplied by the metals. Ionized molecules 
do not play an important role in the EOS or the formation of the spectrum for 
the models we consider here. 

The effect of the gravity on the model structure is large, as expected. In 
particular, the structure of the hotter models ($\Teff \ge 3000\,$K)
with low gravity ($\log(g)\le4$) and close to solar abundances 
($[M/H]\ge -0.5$) is significantly different than the structure of the other 
models. An example is the $\Teff=3500\,$K, $\log(g)=4.0$, $[M/H]=0.0$ model 
shown in Fig.~\ref{struc2}. This model has much higher electron temperatures 
and lower pressures than the corresponding model for $\log(g)=5.0$. This is 
caused by less-efficient convective energy transport in the $\log(g)=4.0$ 
model as compared to the $\log(g)=5.0$ model, which increases the temperature
gradient. 

\subsection{The synthetic spectra}

Fig.~\ref{VB10} illustrates how molecular opacities dominate over the 
spectral range through which M~dwarfs radiate the bulk of their emergent 
flux ($\sim 1{\mu}m$), leaving practically {\em no} window of true continuum. 
The ${\rm H}^-$ continuum lies more than 30\% above the spectrum of the star,
and the TiO absorption bands block more than 45\% of the flux below $1{\mu}m$. 
The flux distributions are very different from the flux distributions of 
blackbodies of corresponding effective temperatures such that the blackbody 
distribution is a poor approximation, even redward of $2{\mu}m$.  

In this figure, we have also plotted the observed energy distribution
of the dM8 star VB$\,$10 for comparison.  The model spectrum (dotted line) 
fits reasonably well the near-infrared spectrum and the depth of the CO bands 
near 2.3 ${\mu}m$, but fails to reproduce the water bands in the IR. It is
not yet clear to what extent the flux calibration of the observed spectrum
around $1.4 \mu$m is responsible for part of this discrepancy. But a large
share could possibly be attributed to shortcomings in the Ludwig \water\ 
opacities.  Schryber \etal\ (1994)\nocite{schryb94} have recently compared 
the Ludwig opacities to other sources and found systematic errors in the 
former. In particular, they concluded that the Ludwig values overestimate 
the opacity of water by a factor of 2-3 in the 1 to 2 $\mu$m spectral range
at the relevant temperatures for M~dwarfs atmospheres. However, these 
conclusions are based on comparisons with other water opacity sources which
are inadequate in this temperature range. The identification of the sources
of error at play here and their solution await better water-line lists 
valid for high temperatures.

\subsubsection{Spectral sensitivity to effective temperature}

The energy distributions of late-type dwarfs are very peculiar and in some 
ways, counter-intuitive.  For example, the molecular opacities which globally 
define the continuum lock in place the peak wavelength at around 1.1~$\mu$m 
for solar metallicities.  As $\teff$ decreases, the peak doesn't move redward 
as one would expect for a blackbody. This is shown in Fig.~\ref{grteff} 
where the strong dependence of the TiO (optical) and water (IR) bands is 
clearly apparent.  Therefore, although the stars do get redder, it is more 
due to a decrease in the visual flux than an increase in red.  When looking 
for photometric bands most sensitive to very cool objects like brown dwarfs, 
it is vital to keep this in mind.  The bandpasses where most of the flux 
escapes are (in order of decreasing importance) $J$, $H$ and $K$. Beyond the 
$K$ band, the flux drops precipitously into a series of water bands growing in 
strength while the emergent flux decreases.

The dependence of the various spectral features on the effective temperature 
is demonstrated in Fig.~\ref{tseq}. These spectra have been folded with a 
Gaussian kernel to correspond to a resolution of $18\ang$ (the 
resolution of the observations by \cite{kirk93}). We have plotted 5 models 
with $\log(g)=5.0$ and solar abundances but with $\Teff$ ranging from 
$4000\,$K (full curve) down to $2000\,$K (dotted curve). 

The important bands of the CaH molecule near $6200\ang$, $6400\ang$ and 
$6850\ang$ number among the most gravity-sensitive features of M~dwarf 
spectra but are unfortunately completely blended with less sensitive TiO 
and VO features. The TiO bands and the CO first overtone are excellent 
$\Teff$ indicators, and should therefore be used to estimate $\Teff$ 
from observed spectra. The atomic lines typically get stronger and 
larger numbers of them succeed in piercing the molecular continuum with 
increasing effective temperatures.  At the very low $\Teff$, only a few 
of the strongest resonance lines are still visible in the spectrum.

\subsubsection{Effects of metallicity and gravity on the spectra}

In the coolest metal-poor objects, the CIA H$_2$ opacities, which are centered 
on 2~$\mu$m, further depress the continuum such that the flux emerges only 
in passbands {\it bluer} than $H$.  This is depicted in Fig.~\ref{grmet} 
where the IR flux distributions of three typical models are shown: one for 
metal-rich conditions expected in younger brown dwarfs bright enough to be 
seen; the others for halo and Pop. III stars.  This figure shows clearly how 
the maximum of the energy distribution is progressively shifted to the 
{\em blue}. This is due to the progressive disappearance of the TiO 
molecule which dominates the opacity below $1.1\,\mu$m in the models of 
higher metallicities: the TiO absorption gives way to the H$^-$ continuum 
and the radiation can escape from deeper, hence hotter, layers of the 
atmosphere. Moreover, a weakening of the H$^-$ opacities themselves, owing 
to the rapidly decreasing availability of free electrons in the metal-poor 
regimes, enhances this effect.  This combines with the growing wide CIA 
band opacities centred near 2~$\mu$m to dramatically increase the visual 
flux and at the same time, decrease the IR flux. This change of 
the opacity as a function of wavelength also causes the effect -- 
easily seen in the synthetic photometry presented below -- that the 
metal-poor models actually appear bluer and hotter than metal-rich models.

Figure~\ref{lte30mh} shows the effects of the metallicity on the spectral
features of the synthetic spectra with $\Teff=3000\,$K and $\log(g)=5.0$ in 
the optical (panel a) and the IR (panel b). The strength of the optical TiO 
and IR water \& CO bands is reduced drastically, whereas the relative strength 
of the metal hydrides (preserved with higher probability than molecules 
involving metals only) increases. This is seen in particular in the bands 
of the MgH, CaH, and FeH molecules. This effect is also apparent in the 
strengths of atomic lines which also become stronger as metallicity decreases.  
This results from the increased gas pressures of metal-poor photospheres 
which compensate for the reduced number densities of atoms in the gas.  With 
the exception of a few resonance transitions, most atomic lines finally 
disappear in the lowest metallicity models.

Panels a) to c) of Fig.~\ref{maxener} show the gravity sensitivity of the 
synthetic spectra around the maximum of its energy distribution 
(0.9--$1.4\,\mu$m) for the three representative metallicities. Since 
the issue of the gravity sensitivity of cool dwarfs energy distributions is 
particularly relevant to the search for spectral signatures of brown dwarfs, 
we have chosen sequences with $\Teff=1600\,$K to illustrate the subject. 

While the overall energy distributions of metal-rich models show little 
dependence on gravity, measurable effects are apparent in the relative 
strength of molecular bands: hydride bands (for reasons mentioned above) 
show a stronger pressure sensitivity than other metallic compounds. However, 
this is often masked by the simultaneous dependence of the band strength on 
the metallicity. Gravity effects can also be detected in the wings of strong 
atomic lines. Therefore, it seems more accurate to derive the gravity 
from line-width analyses rather than from low-resolution spectra. 

The highly non-linear coupling between the parameters $\Teff$, $\log(g)$
and $[M/H]$ introduces the problem that changes in one of the parameters
can be (at least partly) offset by changes in the other two parameters. 
For example, a change in $\Teff$ can be partially simulated by a change
in metallicity (higher $\Teff$ corresponds roughly to lower $[M/H]$). 
However, this is only true at low resolutions; the potential 
ambiguity is removed by combining low-resolution with high-resolution 
spectra in some wavelength bands owing to the fact that line profiles 
react differently to changes of the parameters than the overall energy 
distribution. Therefore, a spectral analysis of M~dwarfs should always 
use both low-resolution and high-resolution spectra in order to derive 
parameters with good accuracy. 

We demonstrate this in Fig.~\ref{irhugh} by showing high-resolution infrared 
spectra between $1.14\,\mu$m and $1.24\,\mu$m for a number of different 
effective temperatures and gravities for solar abundances. We have chosen 
this particular range because high-resolution observations of M~dwarfs are 
available and the synthetic spectra are currently being used to analyze these 
data (Jones 1994, in preparation). In each panel, the model with the highest 
$\log(g)$ also has the greatest line width. For the models with 
$\Teff \le 2000\,$K, the gravity changes not only the line width but also 
strongly influences the strength of several FeH bands found in this 
spectral region. For higher $\Teff$, the influence of molecular opacity 
is much smaller and at $\Teff=4000\,$K, the spectrum is dominated by metal 
lines. The width of these lines is of course very gravity dependent and can be
used to obtain the gravity from analyses of the observed spectra in this range.

\subsubsection{Spectral signatures of brown dwarfs}

According to evolutionary models, M dwarfs and substellar objects can share 
similar atmospheric properties.  Young brown dwarfs pass through a 
deuterium-burning phase on their cooling tracks which heats the atmosphere 
through efficient convective mixing and mimics higher-mass M dwarfs.  A  
potential way to identify brown dwarfs unambiguously was proposed by Rebolo 
\etal\ (1992)\nocite{rebolo92}: the detection of the Li~I resonance doublet at $6780$\AA.  This 
method is based on the principle that the atmospheres of cool dwarfs are fully 
convective up to very low optical depths.  Highly efficient mixing takes atomic
lithium into the hot interior where it is destroyed {\it unless} the central 
temperature does not reach the Li-burning limit, which should be the case 
for brown dwarfs of $M < 0.06 M_{\odot}$ according to interior models. 

The formation of atomic lines like the Li doublet in atmospheres of cool 
dwarfs depends, beyond the abundance, on various factors: molecular 
binding, departures from LTE, treatment of convection and even feedback 
of chromospheric radiation. Pavlenko \etal\ (1994)\nocite{pav94} have superficially 
investigated these effects using their spectral synthesis code (which is 
based on parts of ATLAS) combined with independent model structures (present
work).  To explore these factors with an atmospheric structure which adjusts 
self-consistently to the assumed conditions, we have conducted a separate 
analysis using our code \phx.  As a first step, we generated synthetic 
profiles of the Li doublet in a very cool model to look for molecular 
binding of Li assuming LTE. This is probably not a good assumption, even 
for $\Teff$ as low as $1800\,$K; however, it can be used as a rough guideline 
for the formation of the doublet. (We are currently working on an NLTE 
treatment for Li~I.) 

In Fig.~\ref{lithium} we show synthetic line profiles of the Li~I 
lines for a model with $\Teff=1800\,$K, $\log(g)=4.5$ and solar abundances 
(with the exception of lithium). We vary the lithium abundance from 
$\log\eps_{\rm Li}=3.31$ (the meteoritic abundance of Li, dashed curve in 
Fig~\ref{lithium}) down to $\log\eps_{\rm Li}=0.0$ (topmost curve). The dotted 
curve corresponds to solar Li abundance ($\log\eps_{\rm Li}=1.16$). The differences
in the profiles between the solar and meteoritic abundances are large, even in 
the line wings which are typically not very sensitive to NLTE effects.  This 
shows that the Li~I lines can be used for low $\Teff$ to distinguish brown 
dwarfs from M dwarfs, even though most of the Li nuclei are bound in molecules
like LiCl. For more quantitative analyses of the Li abundance, NLTE models 
must be used. A major problem with NLTE calculations at such low electron 
temperatures are the collisional rates. Collisions with electrons are 
suppressed by the very low number density of electrons in the atmosphere,
however, the collisional cross-section of collisions between atoms and the 
most important species (H$_2$) are poorly known. Therefore, NLTE line profile 
will depend sensitively on the particular assumptions of the collisional cross
sections. 

Significantly, none of the existing brown dwarf candidates could be confirmed 
by Marcy \etal\ (1994)\nocite{marcy94} and Martin \etal\ (1994)\nocite{martin94}, who set very low limits on 
the Li~I doublet strength using new high-resolution high-S/N spectra.  Still, 
this does not rule out the existence of brown dwarfs, nor does it establish 
that the candidates are truly stellar since they could be simply transition 
objects just massive enough to destroy Li without ever reaching a state of 
stable hydrogen-burning.  It is clear that other {\sl independent} spectral 
signatures are needed.  The next generation of models -- incorporating 
line-by-line sampling of molecular opacities -- may also reveal global 
$\log g$ effects as yet unrecognised. 


\subsection{Colors for M~(sub)dwarfs}\label{colors}

\subsubsection{Synthetic photometry}

In order to transform our synthetic spectra into photometric colors that are 
readily applicable to multicolor observations in the optical and infrared, we 
have computed synthetic VRIJHKLL'M magnitudes using our model spectra. The 
synthetic colors were computed by integrating the model fluxes over each 
bandpass.  We adopted the following filter responses from the literature and 
other sources:

\begin{itemize}
\item V, R, I (Cousin) system from Bessell (1990)\nocite{bes90} 
\item J, H, K CIT filters, accounting for atmospheric 
          transmission (Persson, priv. communication). This system
          was originally defined by Elias (1982)\nocite{elias82}
	  and is particularly relevant to the large JHK survey 
          of M dwarfs by Leggett (1992)\nocite{leggett92}. (JHK 
          photometry on the Glass-Johnson system of filters described by 
          Bessell \& Brett (1988)\nocite{bes88} has also been computed.)
\item L, L' (AAO), and M, as described by Bessell \& Brett (1988)\nocite{bes88}
\end{itemize}

To calibrate the flux-magnitude scale, we used the energy distribution of 
Vega as observed by Hayes (1985)\nocite{hayes85} and  Hayes \& Latham (1975)
\nocite{hayes75} in the optical, and by Mountain et al (1985)\nocite{mount85}
in the IR.  Zero magnitudes (including V) and, correspondingly, zero color 
indices were assumed for Vega. 

There is some debate about how to define the zero-point of the magnitude 
scale: some people have used a sample of several A0V stars to derive an 
`average' zero-mag definition; others have simply adopted Vega as the 
definition of zero at all wavelengths.  We have adopted the latter approach 
and assumed V = 0 for Vega, even though the currently accepted value is 
V = +0.03.  While this does not affect the synthetic colors, we caution 
that our absolute flux calibration may be somewhat uncertain.

Our zero-point magnitudes and fluxes are summarized in Table~\ref{coltab}. 
These generally agree within 3\% to corresponding values derived for the 
same filters by Bessell and Brett (1988)\nocite{bes88} and Bessell (1990)
\nocite{bes90} (also shown in Table~\ref{coltab}), even though they assumed 
V(Vega) = +0.03 and used a theoretical energy distribution for the star 
(\cite{dreil80}).
In fact, adopting the larger magnitude {\em and} theoretical fluxes is 
essentially equivalent to assuming V = 0 and adopting the slightly lower 
fluxes published by Hayes and Latham (1975)\nocite{hayes75}.  As a result, 
most of the differences between our approach and that of Bessell and Brett 
tend to cancel out.  We also present in Table~\ref{coltab} the effective 
wavelength for each bandpass for both the observed spectrum of Vega 
($F_{\lambda}$) and for a typical dM8 star based on the best fitting model 
of our grid to VB10: $T_{\rm eff} = 2800\,$K, $\log(g) = 5.0$ and [M/H] =
0.0.  We show the resulting relation between different color indices and 
the effective temperature in Figs.~\ref{ccvik} and \ref{ccjhk}.  In addition, we have 
calculated bolometric corrections for the V magnitude using the filter 
described above. For calibration, we have used a solar model atmosphere 
constructed with \phx. The colors of our solar model agree within reasonable
expectations with the observed colors so that this approach seems acceptable. 

The tables giving photometric colors and bolometric corrections are not 
included in the paper in the interest of brevity, but are available in 
electronic form from the authors upon request.

\subsubsection{Comparison to observations}

One of the largest homogeneous samples of optical-IR photometry of 
red dwarfs has been compiled by Leggett (1992)\nocite{leggett92}, 
who obtained CIT photometry for a set of 322 stars.  This provides 
an excellent basis of comparison for our synthetic colors, which we 
present in the form of color-color diagrams in Figs.~\ref{ccvik} and
\ref{ccjhk}. In general, the agreement between the synthetic colors 
and the observations is good in various color-color planes; in fact 
it is the best that has been achieved so far. Here we present only 
selected diagrams for the sake of brevity. 

The I-K vs V-K diagram is a good metallicity discriminant for the 
coolest dwarfs. As metallicity decreases the models become decreasingly 
sensitive to $\Teff$. When the metallicity reaches typical values for 
halo populations ([M/H]${\approx}-2.0$), I-K is virtually insensitive 
to $\Teff$. This results from a competition between the CIA opacities 
and the flux redistribution in the K band. Beyond the [M/H]${\approx}-2.0$
turning point, the CIA wins and drags the I-K to bluer values with 
decreasing $\Teff$. Note that the $\Teff$ sequences shrink globally in
both colors in Fig.\ref{ccvik}: while the metal-rich stars of Leggett's 
sample reach only temperatures as low as 2300$\,$K, the presented metal-poor 
sequences can reach easily $\Teff$ values of 1500$\,$K within that window.

All curves meet at the hotter end of the sequence ($\Teff>3700\,$K). This 
indicates the diminished impact of molecular opacities at those temperatures. 
There, the models appear to meet the observations perfectly. The models 
indicate that most of the stars in Leggett's sample have metallicities 
from solar to $-1.0$, which could in principle be consistent with their 
membership to the disc population as inferred from space motion analysis 
(Meusinger 1991; Carlberg \etal\ 1985).
\nocite{meus91,carl85} However, the relative rarity of 
solar metallicity stars among the reddest part of the sample appears 
unnatural. We rather believe that this could be due to our use of the 
straight mean assumption for the main opacities: even lower metallicities 
toward the cooler regimes are required in order to 
compensate for an overestimation of the overall opacities in the models.
This behavior resumes when saturation occurs in the molecular bands (here 
principally \water), as Fig.~\ref{ccvik} suggests for the reddest
stars.

This problem is also apparent in the JHK diagram, where the models are 
systematically bluer in J-H than the bulk of the observed colors by at 
least 0.1 mag. Here the situation is particularly complex. Although 
the stars are spread widely through this plane, the models suggest that 
interpretation of their atmospheric parameters based on JHK colors alone 
would not be straightforward. In Fig.\ref{ccjhk}, we have reproduced the 
$\Teff$ sequences 
for solar metallicity, a metallicity of [M/H]$=-0.5$ and several metal-poor 
sequences.  As can be seen, the metal-rich sequences curl around from the 
[blue H-K, red J-H] end (hot end) to the [red H-K, blue J-H] corner 
(cool end), crossing on the way the lower portion of the diagram. This 
behavior reflects once again the struggle of the stars to release their 
fluxes as $\Teff$ decreases: the growing IR molecular opacities efficiently 
redden the energy distribution until saturation in the water bands 
and redistribution effects finally stabilize the emergent fluxes in the 
water opacity troughs where the H and K bandpasses sit. If water opacities
are overestimated, either due to the SM approximation or because of 
independent shortcomings of the Ludwig data, the consequences would be 
simply to exaggerate the amplitude of these loops in the metal-rich regime
where water still dominates the IR energy distribution. These effects are
clearly apparent in Fig.~\ref{ccjhk}. The increasing action of the CIA 
opacities is also seen in this JHK diagram. Here the CIA opacities cause 
the loops to curl back on themselves by [M/H]$=-0.5$. The halo 
population is expected to fall along nearly straight lines toward the blue 
end of this plane.  

The discrepancies noted above are least apparent in color diagrams involving 
on both axes a red-IR color: both axes are than shifted blueward by a similar 
amount. As the TiO and \water\ opacities give way to continuous opacities 
toward the low metallicity regime, the present models provide a better 
representation of low metallicities and halo stars. In order to resolve 
the remaining discrepancies, models with a full line-by-line treatment of 
the water and TiO opacity will be required. 

\section{Conclusions}

In this paper we have described the results of our latest model 
atmospheres and synthetic spectra for M (sub)dwarfs, based on 
straight mean opacities for the dominant molecular absorbers, i.e. 
\water\ and TiO. This new model grid constitutes a significant 
improvement over the earlier models by Allard (1990)\nocite{allard90}. 
We have introduced an opacity sampling treatment of atomic and 
molecular lines and included the FeH bands as well as several new
molecular opacities. 

We have computed an extensive 'base model grid' of more than 700 
cool dwarf atmospheres.  Despite several discrepancies noted in the
presented comparisons with observations, the new model spectra better 
define the basic properties of M~(sub)dwarf radiation fields.  
The synthetic spectra demonstrate the influence of the metallicity 
and gravity on low-resolution spectra.  While we found that  
gravity has a relatively small influence on the low-resolution spectra 
and the overall energy distribution, it has a significant effect on 
high-resolution line profiles and details of the band-systems. However, 
the effects of the gravity become stronger with lower effective 
temperatures. The metallicity has, on the other hand, a large effect 
on both the low and high-resolution spectra. We found that the parameters 
of an observed star are probably best determined by using a combination 
of low-resolution and high-resolution spectra in order to minimize 
ambiguity generated by the coupling between the three parameters $\Teff$, 
$\log(g)$ and $[M/H]$. 

We have also computed synthetic colors for various bandpasses based 
on our low-resolution spectra. The synthetic colors agree now within 
reasonable expectations with observations. The color-color diagrams 
and spectra offer much-needed alternatives to blackbody fluxes 
for the classification of very low mass stars and the estimation of 
their stellar parameters. 

One serious problem which our models face remains the 
predicted strength of the IR water band, which leads to unreliable 
predictions in the JHK color plane. This difficulty will not be
resolved until new more appropriate \water\ opacities become available.

Other improvements to the models immediately suggest themselves.
For example, we have neglected NLTE effects. This may not 
be a good assumption for treating individual metal lines. The 
collisional rates could be quite small owing to the lack of free 
electrons and the relatively small rates for collisions with, e.g., 
H and H$_2$. We are currently working on an extension of our handling 
of collisional processes  
toward lower electron temperatures, which we hope may shed some light 
on the importance of these processes in cool dwarf atmospheres. 

We have also neglected possible external radiation fields. These 
may be important for the chemistry of the upper photospheric layers 
in some instances where non-negligible UV radiation is incident on 
the photosphere from, say, a hotter companion or its own chromosphere.  
In such cases, \phx\ can actually allow self-consistently for the 
effects of external radiation fields on the model atmosphere and on 
the synthetic spectrum (cf.\ \cite{sn93jpap}). We intend to investigate 
these effects in subsequent work.

Finally, the incorporation of more complete and better molecular 
line lists (in particular for water and TiO) and of grain opacities 
remains the first priority in the development of cool dwarf model 
atmospheres. We are presently investigating the effects on the 
models of treating line opacities also for the dominant absorbers 
((\cite{h2olet}, Allard \etal, in preparation).  

The present grid represents our foundation for future improvements. New 
grids are therefore expected to become available much more frequently 
in the future.  Low-resolution synthetic spectra 
and selected high-resolution IR spectra are already available for each 
model atmosphere of this ``base'' grid.  All results can be obtained in 
electronic form upon request to either of the authors.

\acknowledgments

We would like to give special thanks to H. Jones and S. Leggett for 
instructive discussions and for providing their observations in an 
electronic form. We thank the referee for very helpful comments on 
an earlier draft of this paper. We also want to thank S. Starrfield 
for his support and J.M. Matthews for careful reading of the paper. 
Some of the calculations presented in this paper were performed on 
the Cray C90 of the San Diego Supercomputer Center (SDSC); we thank 
them for a generous allocation of computer time.

This work was funded in part by grants to G.~F.~Fahlman and H.~B.~Richer 
from NSERC (Canada); and by a NASA LTSA grant to Arizona State University. 
The development of \phx's opacity database is sponsored by a grant from 
the American Astronomical Society.

\bibliography{yeti,opacity,mdwarf}

\clearpage

\section{Figure captions}

\begin{figure}
\caption{\label{opspc} Molecular opacities dominate over the complete 
spectral range through which M~dwarfs radiate the bulk of their emergent 
flux  ($\sim 1{\mu}m$).} In this figure, the topmost solid curve is
the total opacity distribution included in the present model calculations. 
The opacity sources represented by full lines were used as 
``continuous''-opacities in the solution of the hydrostatic equation near 
$\lambda=1.2{\mu}m$. The dotted curves show some other important molecular 
absorbers (TiO, CN) included in the model spectra.  The total scattering 
(Thomson+Rayleigh H,${\rm H}_2$,He) is also shown as a dashed curve. Panel 
(a) depicts the situation at solar metallicity; panel (b) the corresponding 
situation in the low metallicity regime. Note how the CIA gradually replaces 
\water\ to form the IR coninuum at low metallicities. The $\sim 1{\mu}m$ 
region is governed by VO and FeH opacities.
\end{figure}

\begin{figure}
\caption{\label{VB10} Our best fit to the spectrum of VB$\,$10, based 
on the TiO and CO band strengths.}  (The derived atmospheric parameters are 
$\teff=2800$K, $\log g=5.0$, assuming solar metallicity.)  The observed 
spectrum is from (\cite{kirk93}) ($0.6-0.9\,\mu$m) and Jones (private
communication) 
($0.9-2.5\,\mu$m).  Also shown are the corresponding H$^-$ continuum, 
obtained by neglecting molecular opacities only in the radiative transfer, 
and a Planck distribution of same $\teff$. Notice how the blackbody curve 
still underestimates the K band flux (near 2.2~$\mu$m) even though it is 
depressed by the H$_2$O opacities.
\end{figure}

\begin{figure}
\caption{\label{struc1} The structure of model atmospheres with 
$\Teff=1600\,$K.} Panel (a) gives the run of the gas pressure $P_g$ 
(in dyn/cm$^2$) versus the electron temperature $T_e$ while panel 
(b) gives  the run of the electron pressure $P_e$ (in dyn/cm$^2$) 
versus $T_e$. The dotted lines are for models with $[M/H]=-4.0$ 
whereas the full curves are for models with solar abundances,
$[M/H]=-0.0$. We plot the structure for 3 values of $\log(g)$, 
4.0, 5.0 and 5.0 for both abundances sets. In addition, we show 
the results for $\log(g)=3.5$ for a model with solar abundances. 
The gravity decreases from the top to the bottom curve in each 
abundance set.
\end{figure}

\begin{figure}
\caption{\label{struc2} The structure of a small subset of the model 
atmospheres.} Panel (a) gives the run of the gas pressure $P_g$ (in 
dyn/cm$^2$) versus the electron temperature $T_e$ while panel (b) 
gives the run of the electron pressure $P_e$ (in dyn/cm$^2$) versus 
$T_e$. The dotted lines are for models with $[M/H]=-4.0$ whereas the 
full curves are for models with solar abundances, $[M/H]=-0.0$. We 
plot the structure for 3 values of $\log(g)$, 4.0, 5.0 and 5.5 for 
both abundances sets. The gravity decreases from the top to the bottom 
curve in each abundance set.
\end{figure}

\begin{figure}
\caption{\label{grteff}  Low-resolution synthetic spectra for a model
sequence with $\log(g)=5.0$ and solar metallicity showing the behavior
of the peak wavelengths toward the brown dwarf regime.}
\end{figure}

\begin{figure}
\caption{\label{tseq} The low-resolution synthetic spectra for a model
sequence with $[M/H]=0.0$ and $\log(g)=5.0$ for a number of effective
temperatures.} We plot the synthetic spectra for 5 values of $\Teff$,
$4000\,$K (full curve), $3500\,$K (log dashed curve)
$3000\,$K (dashed-triple-dotted curve), $2800\,$K (dashed-dotted curve)
$2500\,$K (short-dashed curve), and $2000\,$K (dotted curve). The spectra
have been normalized to equal areas for clarity.
\end{figure}

\begin{figure}
\caption{\label{grmet} Low-resolution synthetic spectra for models of 
different metallicities corresponding roughly to the young disk ([M/H] = 0.0),
halo ([M/H] = $-2.0$), and Pop. III ([M/H] = $-4.0$). The gravity is  
$\log g=5.0$ for all models.}
\end{figure}

\begin{figure}
\caption{\label{lte30mh} The low-resolution synthetic spectra for a model 
sequence with $\Teff=3000\,$K and $\log(g)=5.0$ around the maximum of the 
energy distribution.} 
In each panel, we plot the synthetic spectra for 5 values 
of $[M/H]$, 0.0 (solar abundances, full curve), -1.0 (short dashed curve),
-2.0 (dash-dotted curve), -3.0 (dashed-double dotted curve), and -4.0 (long 
dashed curve). The dotted curve gives the blackbody energy distribution for 
$T=\Teff$.
\end{figure}

\begin{figure}
\caption{\label{maxener} The low-resolution synthetic spectra around the 
maximum of its energy distribution (0.9--$1.4\,\mu$m) for a model
sequence with $\Teff=1600\,$K.} Panels (a) to (c) give the spectra for models 
with metallicities of $[M/H]=0.0$,  $-2.0$, and $-4.0$, respectively. In each 
panel, we plot the synthetic spectra for 5 values of $\log(g)$, 3.5 
(dashed-double dotted curve), 4.0 (short dashed curve), 4.5 (dash-dotted 
curve), 5.0 (full curve) and 5.5 (long dashed curve). The dotted curve gives 
the blackbody energy distribution for $T=1600\,$K.
\end{figure}

\begin{figure}
\caption{\label{irhugh} The high-resolution IR synthetic spectra from
$1.14\,\mu$m to $1.24\,\mu$m  for a model sequence with solar abundances.}
Panels (a) to (e) give the spectra for models  with effective temperatures 
of 1800, 2000, 2500, 2800, and $4000\,$K, respectively. In each panel, we 
plot the synthetic spectra for 3 values of $\log(g)$, 3.5 (dashed-double 
dotted curve), 4.0 (short dashed curve), 4.5 (dash-dotted curve), 5.0 (full 
curve) and 5.5 (long dashed curve). The dotted curve gives the blackbody
energy distribution for $T=\Teff$.
\end{figure}

\begin{figure}
\caption{\label{lithium} LTE synthetic Li~I line profiles for a model with 
$\Teff=1800\,$K, $\log(g)=4.5$ and solar abundances (except Li).} The dashed 
line gives give the line profile for $\eps_{\rm Li}=3.31$ (meteoritic Li 
abundance), the dotted line the line profile for $\eps_{\rm Li}=1.16$ (solar 
photosphere Li abundance). The full curves give the line profiles for 
$\log(N_{\rm Li}) \equiv \eps_{\rm Li}=3.0$, 2.5, 2.0, 1.5, 1.0, and 0.0 
(lower to upper curves, respectively) relative to $\log(N_{\rm H})=12$ by 
number.
\end{figure}

\begin{figure}
\caption{\label{ccvik} Synthetic color indices as function of the effective 
temperature for a range of metallicities and gravities:} solar (long dashed 
curves), [M/H] = $-0.5$ (full lines), [M/H] = $-1.0$ (dot-dashed curves), 
[M/H] = $-1.5$ (short-dash long-dashed curves, and [M/H] = $-2.0$ (short 
dashed curves). The values of gravity considered here are $\log(g)=4.5$ (uppermost 
curve), $\log(g)=5.0$ (middle curve), and $\log(g)=5.5$ (lowest curve) for 
each mixture.  The observations of Leggett (1992)\nocite{leggett92} are 
represented by star symbols. The reddest stars in this diagram, LHS$\,$2397a 
and LHS$\,$2924, are found to be as cool as 2400$\,$K and 2300$\,$K 
respectively, while the bulk of the sample coincides with models of 
$\Teff{\approx}3700\,$K if solar metallicity is assumed.
\end{figure}

\begin{figure}
\caption{\label{ccjhk} Same as Fig.~12 in the JHK color plane.}
Here only two values of the gravities ($\log(g)=4.5$ and $5.0$) are shown 
for [M/H] = $-0.5$ to preserve the clarity of this diagram. The gravity
increases from the outer curve to the inner curve for metal-rich mixtures.
In the metal-poor regime, gravity increases and metallicity decreases
toward bluer H-K values.  The hot portions of the sequences ($>3500\,$K) 
cross around J-H$=0.5$. The cool end extends, for the metal-rich models,
toward the red portion of the diagram; for the metal-poor models, to the
opposite blue portion of the diagram. The absence of stars in the latter
section reflects the nearly solar metallicity of the Leggett sample.   
\end{figure}

\clearpage
\section{tables}
\begin{table}
\caption{\label{tab1}}
\end{table}
\begin{table}
\caption{\label{tab2}}
\end{table}
\clearpage
\begin{table}
\caption{\label{coltab}Effective Wavelengths$^1$, Zeropoint Fluxes$^2$ and
Magnitudes$^3$} 
\begin{tabular}{|crrrrrrrrr|}\hline 
& & & & & & & & & \\
                       &  V~~  &  R~~  &  I~~  &  J~~  &  H~~  &  K~~  &  L~~~ &  L'~~ &  M~~  \\
& & & & & & & & & \\
$\lambda_{\rm eff}$    & 0.544 & 0.641 & 0.798 & 1.221 & 1.631 & 2.153 & 3.452 & 3.802 & 4.754 \\
$\lambda_{\rm eff}^*$  & 0.562 & 0.700 & 0.820 & 1.230 & 1.639 & 2.202 & 3.477 & 3.811 & 4.751 \\
$F_{\lambda}$          &35.808 &30.136 &24.047 &16.188 &10.604 & 6.906 & 3.023 & 2.550 & 1.719 \\
zero-point ZP          & 0.021 & 0.208 & 0.453 & 0.883 & 1.342 & 1.808 & 2.705 & 2.890 & 3.318 \\
Refs. 1 \& 2           & 0.008 & 0.193 & 0.443 & 0.902 & 1.369 & 1.884 & 2.772 & 2.966 & 3.421 \\
  & \multicolumn{9}{c|}{CIT values} \\
$\lambda_{\rm eff}$    &       &       &       & 1.251 & 1.623 & 2.203 &       &       &       \\
$\lambda_{\rm eff}^*$  &       &       &       & 1.256 & 1.637 & 2.214 &       &       &       \\
$F_{\lambda}$          &       &       &       &15.603 &10.636 & 6.641 &       &       &       \\
zero-point ZP          &       &       &       & 0.923 & 1.339 & 1.850 &       &       &       \\ 
& & & & & & & & & \\
\hline
\end{tabular}

$^1$Wavelengths in $\mu$

$^2F_{\lambda}$ is $10^{-21}$ erg/sec/cm$^2$/\AA) for a 0.00 magnitude star in each 
   adopted passband

$^3$Mag $= -2.5 {\rm log}(F_{\lambda}) - 48.594 - ZP$

$^*$Based on the best fitting model to VB10 ($T_{\rm eff} = 2800$K; ${\rm log}g = 5.0$, 
solar composition)
 
\end{table}
\end{document}